\documentclass[aps,amsmath,amssymb,nofootinbib,superscriptaddress,showpacs,floatfix,prl,twocolumn]{revtex4-1}

\usepackage{latexsym}
\usepackage{graphicx}
\usepackage{times,psfrag,subfigure}
\usepackage{enumerate}
\usepackage{amsmath}
\usepackage{dsfont}
\usepackage{dcolumn}
\usepackage{bm,bbm}
\usepackage{sidecap}
\usepackage{color}
\usepackage{latexsym,amsmath,amssymb,bm,euscript}
\usepackage{dsfont}
\usepackage{textcomp}
\usepackage{tabularx}
\usepackage{setspace}
\usepackage{ctable}
\usepackage{sidecap}
\usepackage{placeins}
\usepackage{threeparttable}
\usepackage{multirow}
\usepackage{mathtools}
\usepackage{hyperref}

\def\bk{{\sf k}}

\def\bq{{\sf q}}
\def\bt{{\sf t}}

\def\im{{\rm im}~}

\def\br{{\sf r}}
\def\bQ{{\sf Q}}
\def\bG{{\sf G}}

\def\I{\mathcal{I}}

\def\G{\mathcal{G}}

\def\half{{1\over2}}

\def\bj{{\sf j}}
\def\be{{\sf e}}

\def\bp{{\sf p}}

\def\eig{\ket{\psi^{\nu\bk}}}

\def\eigb{\ket{\psi^{\nu\bk}_b}}
\def\beigb{\bra{\psi^{\nu\bk}_b}}

\def\Tr{\mathop{\mathrm{Tr}}}

\newcommand{\beq}{\begin{equation}}
\newcommand{\eeq}{\end{equation}}
\newcommand{\beqarray}{\begin{eqnarray}}
\newcommand{\eeqarray}{\end{eqnarray}}

\newcommand{\ket}[1]{|#1\rangle}
\newcommand{\bra}[1]{\langle #1|}

\allowdisplaybreaks

\begin{document}

\title{Selection rules for quasiparticle interference with internal nonsymmorphic symmetries}

\date{\today}

\author{Raquel Queiroz}
\email{raquel.queiroz@weizmann.ac.il}
\affiliation{Department of Condensed Matter Physics,
Weizmann Institute of Science,
Rehovot 7610001, Israel}

\author{Ady Stern}
\email{adiel.stern@weizmann.ac.il}
\affiliation{Department of Condensed Matter Physics,
Weizmann Institute of Science,
Rehovot 7610001, Israel}

\begin{abstract}

 We study how nonsymmorphic symmetries that commute with lattice translations are reflected in the quasiparticle interference (QPI) maps measured by scanning tunneling microscopy (STM). QPI maps, which result from scattering of Bloch states off impurities, record the interference of incoming and scattered waves as a function of energy and tip's position. Although both the impurity and the tip generically break spatial symmetries, we find that the QPI maps provide universal information on these symmetries. 
The symmetries impose constraints on the relation between various momentum components of the Bloch functions. These relations result in selection rules on certain momentum transfers in QPI maps. We find that universal information is encoded in the absence of QPI signal, or in the relative intensity of its replications. We show examples for one-dimensional chains and an effective model of the layered compound $\rm ZrSiS$. We discuss the implications of our theory in the analysis of observed QPI of the Weyl semimetal $\rm TaAs$.  Our theory is particularly relevant for materials in rod and layer space groups, or when a correlated order parameter, such as antiferromagnetism, enlarges the unit cell.

\end{abstract}

\date{\today}


\maketitle

\emph{Introduction ---}
Symmetry plays a pivotal role in band theory for the determination of the global features of energy bands, even in the absence of details of the microscopic Hamiltonian  \cite{Dresselhaus_Group_2008,Bradley_The_2010,Bradlyn_Topological_2017,kruthoff2017topological,po2017symmetry-based}.
The block diagonalization of the Hilbert space by symmetry representations can lead to stable or enforced band crossings in the Brillouin zone (BZ) \cite{Herring_Accidental_1937,Zak_Band_1982,Young_Dirac_2015,Po_Filling_2016,Alexandradinata_Topological_2016,Bradlyn_Beyond_2016}. These realize exotic relativistic fermions on the lattice that dominate transport properties when close to the Fermi level \cite{Wehling_Dirac_2014} and are currently the target of active research. To verify their stability, a direct experimental inspection of the symmetry character of Bloch electrons is of unquestioned value, but it is not an easy task. It is notably challenging when the symmetries are nonsymmorphic. These 
are unique to crystalline environments, combining a point-group action with fractional lattice translations.
Direct experimental probes, such as tunneling or photoemission spectroscopy, couple either to momentum ($\hat\bp$) or position ($\hat\br$) eigenvalues and therefore cannot preserve nonsymmorphic symmetries. One would naively think that 
consequently their eigenvalues  
cannot be measured by these techniques. In this work, we see that this is indeed not true: There are 
precise symmetry signatures in the response to spectroscopic probes, which  
may be relevant for the measurement of crystalline topological materials, for example in Ref.~\cite{Ma_Experimental_2017}.

Our strategy is to find universal, symmetry-enforced, selection rules in the interference pattern created by elastically scattered electrons due to dilute impurities.
Fourier-transformed STM, or QPI \cite{Crommie_Imaging_1993,Petersen_Direct_1998,Capriotti_Wave_2003,Pereg-Barnea_Theory_2003,Wang_Quasiparticle_2003}, measures fluctuations of the density of states $N(\br)$ through a differential conductance map around an isolated impurity. Through a Fourier transformation, it is possible to identify the contribution of elastic scattering events according to a fixed momentum transfer $N(\bq)$. The relative intensity of the differential conductance peaks depends on unknown details of the impurity and tunneling elements of the measuring tip, obscuring the interpretation of a QPI pattern. Here we offer an argument based solely on symmetry to extract information from relative peak intensities, remarkably evident in the presence of nonsymmorphic symmetries. Simply put,
\begin{align}
N_{\nu'\!-\nu}(\bq)=0,~~~ \text{if  }~\omega^{(\nu'\!-\nu)+
(\beta-\alpha)}\neq 1\label{sel:nonsymmorphic}. 
\end{align}
where $\omega$ is a phase that characterizes the unitary symmetry, whose eigenvalues at the $\Gamma$-point are  $\omega^\nu$ with $\nu$ an integer; $\alpha$ counts the number of times $\bq$ crosses the BZ boundary along the direction of the fractional translation. QPI replications are labelled by different $\alpha$ along the nonsymmorphic direction. Finally, $\beta$ is determined by the properties of the tunneling tip.  Even though $\beta$ is not generally fixed, a single measurement can strongly favor one. 

Equation \eqref{sel:nonsymmorphic} gives us \emph{a necessary condition for  a $\beta$-tip to observe  QPI amplitude from a transition between two energy eigenstates with symmetry flavors $\nu$ and $\nu'$ through a momentum transfer of  $\bq$, when  $\bq$ crosses the BZ boundary $\alpha$ times}. 
In the following sections we prove this statement by expressing the eigenstates as a  function of the local degrees of freedom, position $\hat\br$ and eigenstates of the point group $\hat R$, resolving how the symmetry is manifested when the system is coupled to a local tunneling tip $\hat M$. We find a decomposition of the energy eigenstates relating the Bloch component and the $\hat R$ eigenvalue, leading to scattering selection rules that depend on $\bq$.

\emph{Eigenstates and internal symmetries ---}
We restrict our study to cases where the symmetry  operator commutes with lattice translations. This implies that it acts on a set of internal degrees of freedom, eigenvalues of a local operator $\hat O$. This operator labels the atomic orbitals, $\phi_o(\br)$ by a quantum number $o$. It can refer to spin or any  
representation of the local point group. The form of 
$\phi_o(\br)$ is determined by microscopic details of the Hamiltonian, but nevertheless highly constrained by the symmetries. An energy eigenstate is  generally written as
\begin{align}
    \textstyle\ket{\psi_b^{\nu\bk}}=\sum_o\int_\br\phi^{\nu\bk, b}_{o}(\br)\ket{o;\br},
\end{align}
where $\nu$ labels the symmetry eigenvalue, $\bk$ the crystal momentum and $b$ a band index, 
omitted in the following. The combined notation $\ket{o;\br}\equiv\ket{o}\otimes\ket{\br}$ stands for eigenstates of both $\hat O$ and  $\hat\br$, forming a basis that satisfies $\bra{o;\br}o';\br'\rangle=\delta_{oo'}\delta(\br-\br')$. 

We consider a discrete transformation, say a rotation,
that respects $\hat R^n=\hat 1$.
Its possible eigenvalues are given by the integer powers of $\omega=\exp\{2\pi i/n\}$, 
with matrix representation $\rho$ in the orbital basis. 
Since $\hat R$ acts on internal degrees of freedom, such as spin or a direction perpendicular to the lattice plane, it acts trivially on the position eigenstates and commutes with lattice translations 
$\hat T_\bj\ket{o;\br}=\ket{o;\br+\bj}$.  The commutation condition is necessary for the validity of Eq.\eqref{sel:nonsymmorphic}.
Examples to such  symmetries are abundant in low dimensional materials or rod and layer space groups.  

While for a symmorphic symmetry $\hat R$ commutes with the Hamiltonian, in a nonsymmorphic symmetry $\hat R$ only commutes with the Hamiltonian when
combined with a fractional translation $\hat T_{\be/n}$ along a unit cell vector $\be$
~\cite{Hiller_Crystallography_1986,Alexandradinata_Topological_2016}. The combined action $\hat R\hat T_{\be/n}$ relates different points in the unit cell creating a Wyckoff orbit of multiplicity $n$ 
that ends 
shifted by a full 
lattice translation, 
 $(\hat R\hat T_{\be/n})^n= \hat T_\be$. This fixes the eigenvalues of the nonsymmorphic symmetry to be  $\omega^\nu e^{i\bk\cdot\be /n}$. Note that $\be$ is not necessarily a basis vector. Finally,
 the energy eigenstates find their coefficients conditioned by lattice translation symmetry and the nonsymmorphic symmetry to satisfy {
\begin{align}
\textstyle\phi^{\nu\bk}_{o}(\br)\!=\!e^{i\bk\cdot\bj}\phi^{\nu\bk}_{o}(\br+\bj)\!=\!\omega^{\nu}e^{i{\bk\cdot\be\over n}}\rho_{oo'}^*\phi^{\nu\bk}_{o'}(\br+\be/n).\label{nonsymmcoeff}
\end{align}
}

\begin{figure}
    \centering
    \includegraphics[width=\columnwidth]{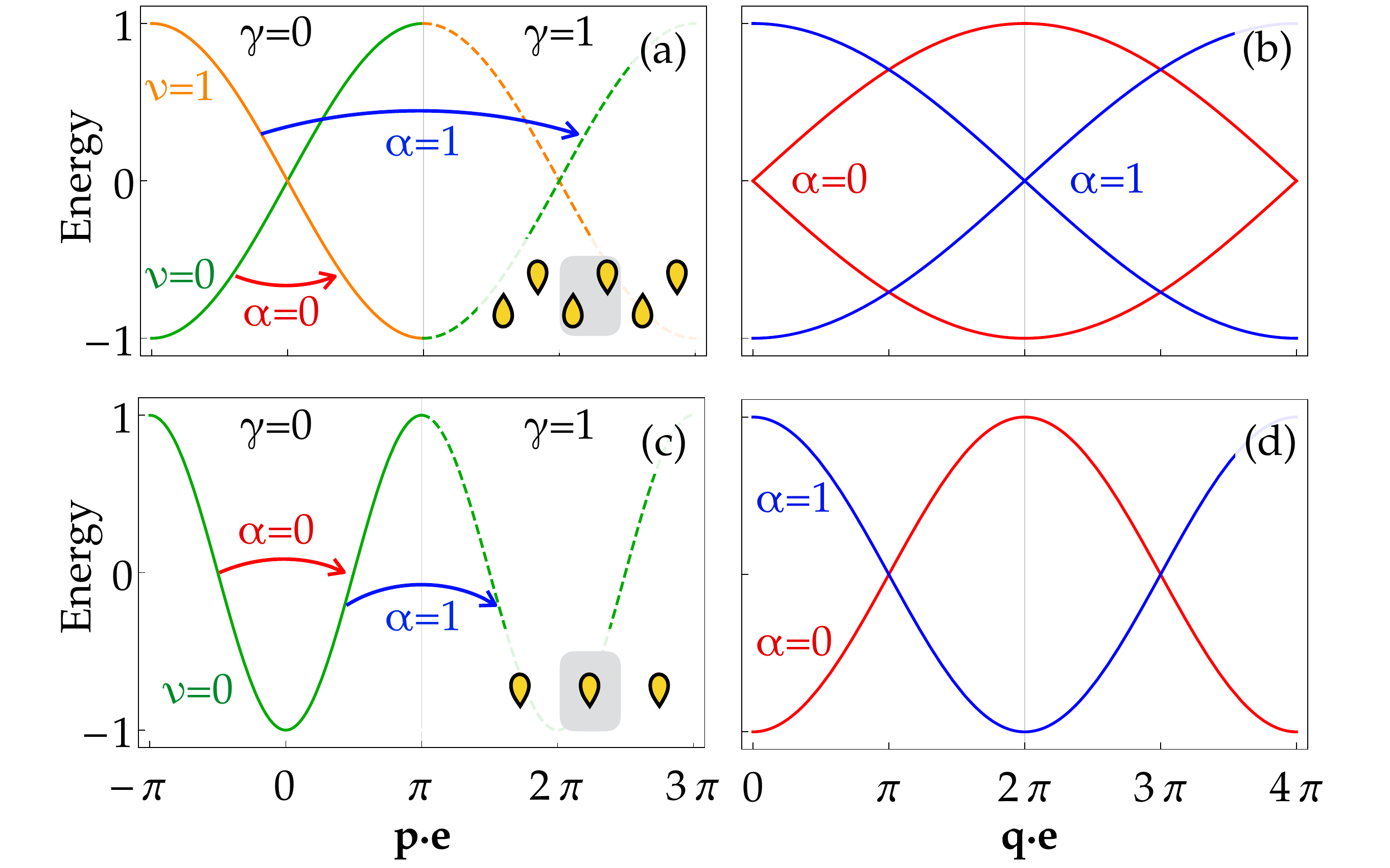}
    \caption{Scheme of the band structure (a,c) and QPI (b,d) of  minimal  nonsymmorphic (a,b) and symmorphic (c,d) systems. The values $\nu$, $\gamma$ and $\alpha$, determine the selection rules imposed by the operator $\hat m$, according to \eqref{sel:nonsymmorphic}: $\nu$ labels the symmetry eigenvalue, $\gamma$ the BZ along the fractional translation $\be$, and $\alpha$ the difference in $\gamma$ of the scattered states. In the nonsymmorphic case, $\alpha$ is selected according to the tip character $\beta$ \eqref{commutation_m}.
    }
    \label{fig:scheme}
\end{figure}

It is illuminating to first see what \eqref{nonsymmcoeff} implies on the form of the energy eigenstates in real and momentum space. Consider $\hat R$ a shift in $o$ by one. We can decompose the position $\br$ to be $\tilde\br-a\be/n-\bj$, where $a=0,...,n-1$. Then $\tilde\br$ is defined in one fraction of the unit cell. Then,
{\begin{align}
  \eig=\sum_{oa\bj}e^{i\bk\cdot({a\over n}\be+\bj)}\omega^{\nu a}\int_{\tilde\br}\phi^{\nu\bk}_o(\tilde\br)\ket{o-a;\tilde\br-{a\over n}\be-\bj},
\end{align}}
Here $\phi_o(\tilde\br)$ is fixed by the details of the Hamiltonian which are not symmetry dictated. The relationship between contributions of different orbital and position eigenstates is, on the other hand, fixed by symmetry. We can look at a simple one-dimensional example with 
$n=2$ and $\rho=\sigma_x$. The band structure, shown Fig.\ref{fig:scheme}(a), corresponds to a hopping tight-binding Hamiltonian  $H(k)=\sin (k/2)~\sigma_x$. With only two bands, there is a perfect locking between the orbital (depicted as upwards and downwards droplets) and the atomic position in the unit cell.

Alternatively, in the eigenbasis of $\hat R$ (with eigenvalue $\omega^\lambda$) and momentum  $\hat\bp=-i\hat \partial_\br$ (with eigenvalue $\bp$), 
Eq. \eqref{nonsymmcoeff} is translated to a restriction on the Fourier coefficients of each Bloch state, 
{ \begin{align}
    \bar\phi^{\nu\bk}_\lambda(\bp)\propto\delta(e^{i\bp\cdot\bj}-e^{i\bk\cdot\bj})\delta(\omega^{\lambda}e^{i{a\over n}(\bp-\bk)\cdot\be}-\omega^\nu).
\end{align}}
The first $\delta$-function implies momentum is fixed to differ from $\bk$ by a reciprocal lattice vector $\bG$ respecting $\exp\{i\bG\cdot\bj\}=1$, which is the essence of Bloch's theorem. Importantly, the second $\delta$-function implies that $\bG$ fixes the eigenvalue of $R$, $\lambda$. In the diagonal basis, $\ket{\lambda;\bp}$ then
\begin{align}
  \textstyle  \eig=\sum_\bG\bar\phi^{\nu\bk}_{\nu-\gamma}(\bk+\bG)\ket{\nu-\gamma;\bk+\bG},\label{nsymeigenstate}
\end{align}
where we define the shorthand  $\gamma\equiv{\bG\cdot\be/ 2\pi}\!\!\!\mod \!n$. 

Note that each Bloch component in \eqref{nsymeigenstate} is characterized by a different point group eigenvalue. From this it follows how bands with different $\nu$'s meet at the boundaries of the BZ: Changing smoothly $\bk\to\bk+\bG\cdot\be$ implies 
 that both $\gamma$ and $\nu$ shift by an integer after the cycle in opposite ways such that $\nu-\gamma$ is unchanged.
That is, there is an adiabatic connection between states of different $\nu$ at the boundaries of the BZ, 
which leads to enforced band crossings  (located at the BZ boundary if time-reversal is present) 
and justifies the large number of degeneracies seen in nonsymmorphic materials~\cite{Zhao_Nonsymmorphic_2016,Young_Dirac_2012,Young_Dirac_2015}.

\emph{Quasiparticle interference ---} QPI
is a direct measurement 
of the elastic scattering  due to dilute impurities.
It is ideal to study symmetry allowed scattering, and it has been extensively used to study 
topological insulators \cite{Zhou_Theory_2009,Guo_Theory_2010,Hofmann_Theory_2013,Roushan_Topological_2009,Beidenkopf_Spatial_2011,Alpichshev_STM_2010}, Graphene \cite{Rutter_Scattering_2007,Pereg-Barnea_Chiral_2008}, and 
high temperature superconductors  \cite{Hoffman_Imaging_2002,Hoffman_A_2002,Pereg-Barnea_Theory_2003,Hanaguri_Unconventional_2010,podolsky2003translational,fujita2014direct}.
An impurity located at $\br_0$ that creates (for example) a Gaussian-shaped potential  $v(\br,\br_0)\!\propto\!\exp\{-\half |\br-\br_0|^2/\xi^2\}$, 
adds the operator
$\hat V\!=\!\int_\br v(\br,\br_0)\ket{\br}\hat v\bra{\br}$, where $\hat v$ acts on the $\hat R$ subspace, to the electron's Hamiltonian. It is generally not diagonal, and assumed to be random. 
The tunneling tip can be similarly described  by  $\hat M(\br)\!=\!\ket{\br}\hat m\bra{\br}$, where $\hat m$ contains the tunneling elements in orbital space. 
 Finally,  in the limit of dilute and weak impurities \cite{Guo_Theory_2010} the measured local density of states is 
captured by $N(\br)=-\im \Lambda(\br)/\pi$  with
\begin{align}
\Lambda(\br)=\Tr \hat M(\br)\mathcal G\hat V\mathcal G,\label{eq:ldos}
\end{align}
where $\G$ is the retarded (unperturbed) Green's function, and the trace is taken over the quantum numbers $\nu$ and $\bk$.
Performing a Fourier transformation we write 
$\hat M(\bq)
=\!\int_\bp\ket\bp\hat m\bra{\bp-\bq}$, and
$\hat V\!=\!\int_{\bp,\bq} v(\bq,\br_0)\ket\bp\hat v\bra{\bp-\bq}$. 
The phase associated with the impurity position remains as an overall prefactor,  
as addressed in Ref.~\cite{torre2016holographic}. As we show in the supplementary material \cite{supp}, the
block diagonalization of the eigenstates
\eqref{nsymeigenstate}, and the consequent diagonalization of $\G$, implies that in Fourier space $\Lambda(\bq)$ can be decomposed in a sum of contributions 
of the form 
\begin{align}\Lambda^{\nu\nu'}_{\bk\bk'\!,\gamma}(\bq,\bQ)=\I^{\nu\nu'}_{\bk\bk'\!,\gamma}(\bq,\bQ)~\hat m^{\nu'\!-\gamma-\alpha~}_{\nu-\gamma}\delta(\bq-\bk'+\bk+\bQ),\label{QPIexpanded}\end{align}
with $\alpha\equiv{\bQ\cdot\be/ 2\pi}\!\!\!\mod \!n$.
Each summand in \eqref{QPIexpanded} describes the interference of incoming and outgoing waves, weighted by a nonuniversal intensity function $\I$. 
The 
lattice periodicity determines that all processes satisfy $\bq-\bk'+\bk= \bQ$, with $\bQ$ a reciprocal lattice vector, fixed by the restriction of $\bk$ and $\bk'$ to the first BZ.
Symmetries within the unit cell impose further conditions, and suppress contributions to the QPI as encoded in the matrix elements of $\hat m$. In the nonsymmorphic case, these suppressed contributions depend on $\bQ$. More precisely they depend on $\alpha$, since it indicates whether momentum is transferred along $\be$, see Fig.~\ref{fig:scheme}. 

The function $\I$, whose explicit form is given in the supplementary material,  includes the nonuniversal details that modulate the intensity of the QPI signal. Those include features associated with the impurity, which is only a tool in the experiment, and features associated with the Bloch coefficients, that are dependent on the non symmetry dictated details of the Hamiltonian. The former 
include the impurity matrix elements, spatial distribution and energy dependence. The exact form of the potential exerted by the impurity may limit the visibility of certain features of the interfering wavefunctions. The latter is determined by the charge distribution within the unit cell. Generally, the first Bloch component is favored and $\alpha=0$ dominates the QPI, but frequently other values of $\alpha$ are observed as well.

\begin{figure}
    \centering
    \includegraphics[width=\columnwidth]{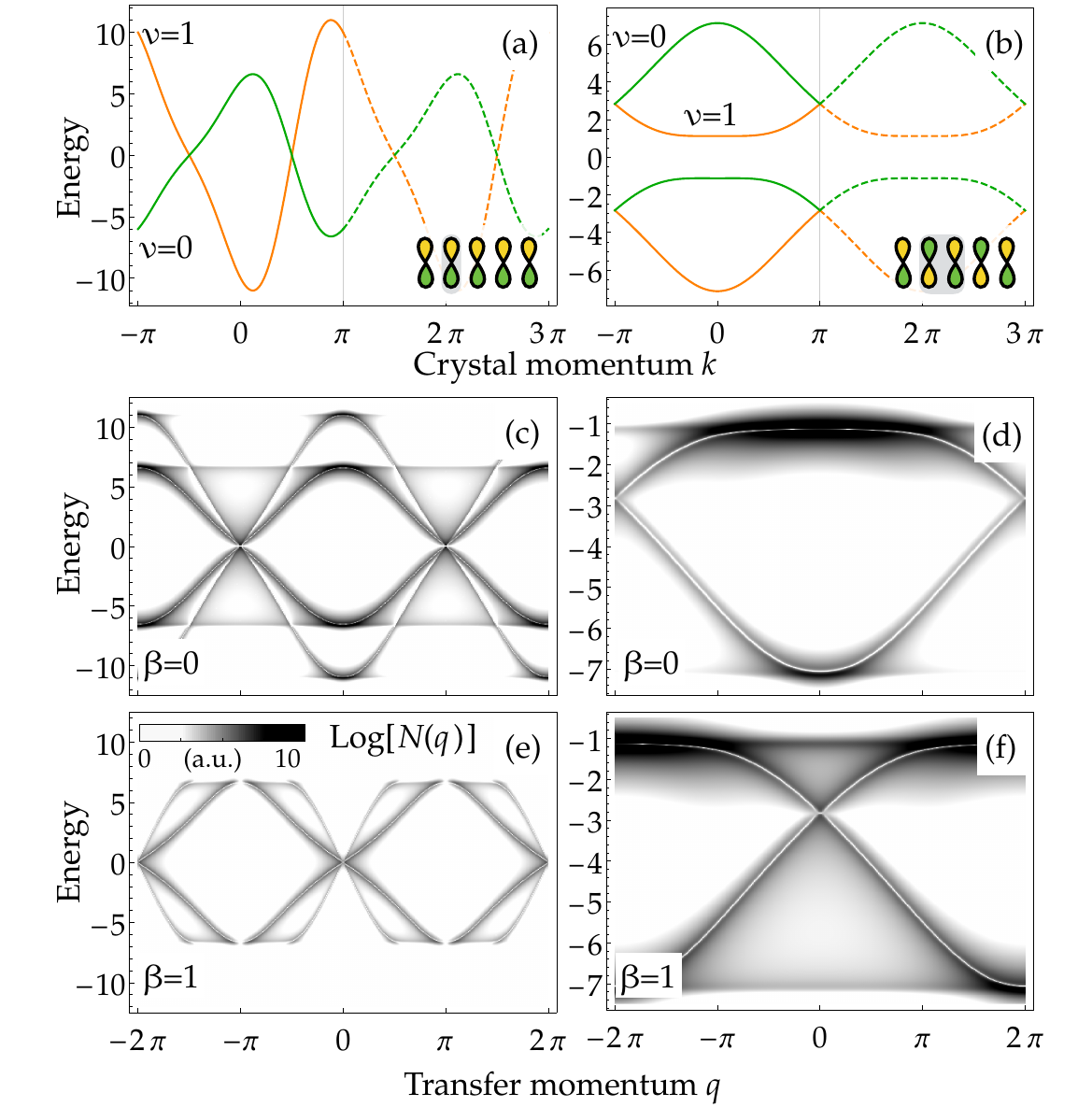}
    \caption{QPI created by a symmorphic (left) and nonsymmorphic (right) one-dimensional semimetal defined in \cite{supp}. (a,b) Band structure, (c-f) QPI, ${\rm Log}|N(q)|$ for two distinct measurements $\beta=0,1$. The QPI obeys the selection rule: (c) Only scattering within same $\nu$, with a $q=\pi$ crossing of QPI replications, no evidence of Dirac point; (e) Only scattering for different $\nu$, green to orange, with $q=0,2\pi$ crossings due to the Dirac points; (d,f) Elastic scattering implies $\nu=\nu'$, thus $\beta$ selects $\alpha=1$ in (d) and $\alpha=0$ in (f). Note that the two replications appear at different $\beta$, and there is no crossing at $\pi$.}
    \label{fig:n2}
\end{figure}

We now focus on what the matrix elements of $\hat m$ can tell us about the symmetry dictated features of the interfering electrons, both in the symmorphic and non-symmorphic cases. In the symmorphic case the matrix element in \eqref{QPIexpanded} 
is substituted by $\hat m^{\nu'}_\nu$. This implies that Eq.\eqref{sel:nonsymmorphic} is satisfied with $\alpha=0$, provided the $\hat m$ respects 
\begin{align}\hat m\hat R=\omega^\beta\hat R\hat m,\label{commutation_m}\end{align}
If, for example, $\hat m$ is diagonal in the eigenspace of $\hat R$, the QPI will manifest only scattering between states of equal 
eigenvalue $\nu$.
This is relevant, for example, in the study of surface states of topological insulators \cite{Guo_Theory_2010,Hofmann_Theory_2013}, where 
$\nu$ relates to spin, and $\beta$ to the polarization of a magnetic tunneling tip.
In 
contrast is the nonsymmorphic case, due to the correlation between the amplitude of the momentum $\bk+\bG$ in the interfering Bloch states 
and the quantum state of the intracell degrees of freedom. The observed channels satisfy Eq.\eqref{sel:nonsymmorphic}, which allows for all scattering channels at all $\beta$. However, different channels are manifested at different replications, that is, at different $\alpha$. In Fig.~\ref{fig:scheme}(b) the two replicated signals will be subject to distinct matrix element effects, and thus strongly vary in intensity.

{Even though a general measurement will consist of a superposition of $\beta$ and will generally not be able to fully suppress replications, the relative intensity of different $\alpha$ contains information about the band representations. There are two ways of retrieving this information. First, in the presence of internal nonsymmorphic symmetries matrix elements will alternate along $\bQ_\be$. If the impurity is sharply localized, allowing for the observation of many replications, the alternating intensity of QPI peaks along a preferred direction is a strong indicator of a nonsymmorphic structure.
Second, $\beta$ can be used as a tuning knob, varying the intensity of different $\alpha$ channels. Experimentally, this can be achieved by combining data of distinct measurements around an impurity, differing in the tip's position within the unit cell. 
Since all Hermitian operators can be decomposed as a sum of operators satisfying \eqref{commutation_m}, combining distinct measurements can be used to isolate $\beta$. 
To clarify the latter proposal consider the idealized example presented in Fig.\ref{fig:n2} (b-inset), with $\rho=\sigma_x$ in the orbital basis. Measuring a single orbital is represented by $  m_\pm=\sigma_0\pm\sigma_z$. 
Then $m_\beta$ can be constructed by $ m_0= m_++{m}_-$ and $m_1= m_+-{m}_-$. Realistically, a fine tuned superposition is needed, but we note that intracell spatial resolution has been successfully used to highlight surface states in {\rm TaAs} \cite{Batabyal_Visualizing_2016,Inoue_Quasiparticle_2016,Gyenis_Imaging_2016}}.

\begin{figure}
    \centering
    \includegraphics[width=
    \columnwidth]{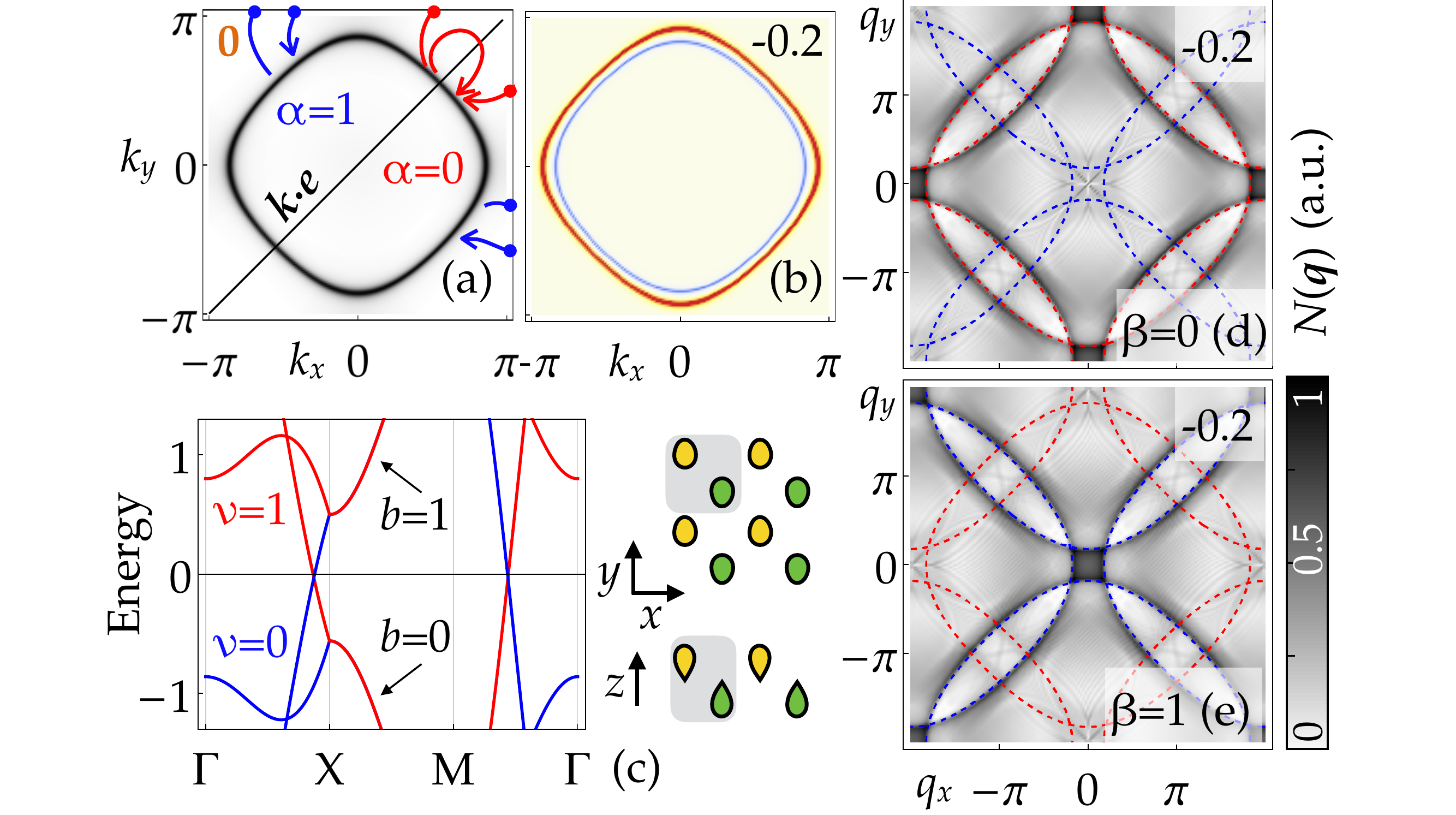}
    \caption{QPI for an effective model of $\rm ZrSiS$, with an inplane glide symmetry:  (a) Density of states at the Fermi level, marked with scattering processes if different $\alpha$; (b) Density of states below the Fermi level, colored by $\nu$. (c) Band structure. 
(d-e) QPI for the $b=0$ band at $E=-0.2$. The dashed lines (blue and red) guide the eye to the contributions from different $\alpha$. The selection rule \eqref{sel:nonsymmorphic} results in the complete suppression of one $\alpha$ QPI channel, and a remarkable qualitative change in the pattern due to the internal glide symmetry.}
    \label{fig:ZrSiS}
\end{figure}

\emph{Numerical simulations ---}
We calculate the QPI numerically for two one-dimensional systems: one symmorphic and one nonsymmorphic and a two-dimensional nonsymmorphic semimetal 
corresponding to an effective model of {\rm ZrSiS}~\cite{Topp_Surface_2017,Schoop_Dirac_2016}. The tight-binding models are given in \cite{supp}. In all cases $\omega=-1$. In three dimensions, these can be applied to rod and layer groups.

First, we look at 
Fig.\ref{fig:n2}. 
In the symmorphic case (panels a, c and e), we show a two-orbital model with first and second neighbour hopping. 
The two bands, which carry different eigenvalues of $\hat R$, cross forming two distinct Dirac points in the BZ. The QPI (panels c and e) depends crucially on $\hat m$ as expected from \eqref{sel:nonsymmorphic} but $\alpha$ plays no role. This implies that the crossing of replications at $q=\pi$ is only present in panel (c), and the Dirac points can only be seen in panel (e) at both $q=0$ and $q=2\pi$. For the nonsymmorphic chain (b, d and f) we consider a four-band model, with two orbitals at each site. We show the QPI for the two lower energy bands.
In contrast with the symmorphic case, here $\beta$ defines which $\alpha$ signal is observed. This is an evidence for the locking between momentum and $\hat R$ eigenvalues. We find in panels (d) and (f) only two lines, and not four. 
The crossing at $q=\pi$ in (d) is absent, and the evidence for the Dirac point 
appears only either at $q=0$ (d) or at $q=2\pi$ (f) but not both. A Dirac point with equal $\nu$ is typical at the boundaries of the BZ in time-reversal symmetric nonsymmorphic systems.
Now we consider the two-dimensional model in Fig.\ref{fig:ZrSiS}.  
It can be generally applied to layered nonsymmorphic materials with a glide plane along the surface, 
with $\be=\hat x+\hat y$.
The relevant physics is well described by a four-band model, with two sets of bands, $b=0,1$, distinguished by a symmorphic symmetry, 
protecting a Dirac ring at the Fermi level.
The nonsymmorphic symmetry 
acts on each set separately. Since we are interested in the scattering selection rules for different $\nu$, we choose to show the QPI of a single set $b=0$. 
Physically, different bands can occupy different regions in the unit cell, and the impurity position can induce scattering primarily in one set. Consequently, the results are blind to 
the Dirac ring. 
We show the density of states (panels a-b) at constant 
energy resolved by $\nu$, and the QPI (d-e) for $\beta=0,1$.
The selection rule \eqref{sel:nonsymmorphic} imposes drastic qualitative differences in the QPI for the different measurements by selecting
different $\alpha$ (red and blue dashed lines). 

 \emph{Application to TaAs --- }
In Refs.\cite{Batabyal_Visualizing_2016,Inoue_Quasiparticle_2016}, Batabyal, \emph{et.al.} and Inoue \emph{et.al.} studied the surface QPI of the Weyl semimetal TaAs, focusing on highlighting its arc states. Apart from arc states, both works study additional states, coined ``bowtie" and ``cigar" surface states, which show $\bQ$ replications. These results are platform to test our predictions. The intensity modulations of the replications are highly anisotropic: They alternate along $\bQ_y$ but not along $\bQ_x$. From our theory, we propose these modulations reveal a (possibly approximate) nonsymmorphic structure of these surface states. Data in Ref.\cite{Inoue_Quasiparticle_2016} further indicates that it predominantly occurs at Ta sites. We have verified this prediction by computing the charge density of the ``bowtie" state, shown in the supplementary material \cite{supp}.
Similar analysis of the intensity of replications can be used to detect an onset of a correlated order-parameter that leads to an increase of the unit cell, such as in antiferromagnets.

\emph{Conclusion ---}
In this letter we propose a set of measurement-based selection rules to explore the symmetry aspects of Bloch electrons with STM. We do so by expressing 
the Green's function in eigenstates of the point-group and momentum operators. When a nonsymmorphic symmetry is present, this decomposition must be performed at each Bloch component of the energy eigenstates
independently.
We show that two factors are crucial to define universal selection rules. First, how the tunneling tip couples with the local orbital degrees of freedom ($\beta$). Second, whether the momentum transfer between energy eigenstates crosses the boundaries of the BZ along the direction of the fractional translation ($\alpha$). The two factors play an analogous role, 
evident in Eq.~\eqref{sel:nonsymmorphic} 
and in the similarity of
Figs.~\ref{fig:n2} (d) and (f). {
Independently of the impurity potential, we show that the relative intensity of QPI replication of different $\alpha$, contains information about the symmetry representation of Bloch bands and can be revealed by data analysis. That is, to find a signature of internal nonsymmorphic symmetries we should compare QPI peaks of different Brillouin zones. We further propose that in order to overcome measuring limitations imposed by the finite extent of the atomic orbitals and impurities, we can alternatively explore the spatial resolution of STM. 
Performing distinct measurements around the same impurity, thereby varying the tunneling elements in orbital space, it is possible to select the tip character $\beta$. If the symmetry is nonsymmorphic the two approaches yield similar results.
}

\begin{acknowledgements}
The authors acknowledge fruitful discussions with A. Rost, L. Muechler, E. Khalaf; N. Morali, B. Yan, N. Avraham and H. Beidenkopf for pointing out unexplained features in the QPI of $\rm TaAs$, and B. Yan and H. Fu for calculating the charge density of its surface states. This work was supported by the Israel Science Foundation; the European Research Council under the Project MUNATOP; the DFG (CRC/Transregio 183, EI 519/7-1).
\end{acknowledgements}

\bibliographystyle{apsrev4-1}

\bibliography{readcube_export.bib}

\clearpage
\onecolumngrid

\section{Supplementary material for \\ Selection rules for quasiparticle interference with internal nonsymmorphic symmetries
}

\section{Derivation of the quasiparticle interference amplitude}

To analyze the quasiparticle interference pattern, as well as other physical responses, it is convenient to block diagonalize the the Green's function with respect to the crystal momentum $\bk$ and symmetry flavor $\nu$,
\begin{align}
    \G_b^{\nu\bk}(E)={\eigb\beigb\over (E-i0^+)-\varepsilon_b^{\nu\bk}}
\end{align}
 Here $\varepsilon_b^{\nu\bk}$ is the band energy, and we omit the explicit dependence in energy, $E$, to avoid cluttering. Using the eigenstate decomposition in the main text, Eq.6, we find,
\begin{align}\mathcal G^{\nu\bk}= \sum_{\bG\bG'}\ket{\nu-\gamma;\bk+\bG} I^{\nu\bk}({\bG,\bG'})\bra{\nu-\gamma';\bk+\bG'}\label{ultimate_block},\end{align}
where $\gamma={m\bG\cdot\be/ 2\pi}\!\!\!\mod \!n$ and $\bG$ is a reciprocal lattice vector. The relative intensity of each Bloch component takes the explicit form 
\begin{align}   I^{\nu\bk}({\bG,\bG'})=\sum_{b}{\bar\phi^b_{\nu-\gamma}(\bk+\bG)\bar\phi^{b}_{\nu-\gamma'}\!\!\!\!\!\!\!\!{}^*~~~ (\bk+\bG')\over (E-i0^+)-\varepsilon_b^{\nu\bk}}\label{intensit}
.\end{align}
While the divergence 
from the denominator only depends on $\bk$, $I^{\nu\bk}(\bG,\bG')$ exponentially decays with $\bG$ as a
consequence of the  
spatial width of the atomic wavefunctions. 

We calculate the quasiparticle interference amplitude by Fourier transforming the local density of states, which is given by
\begin{align}
N(\br)=-{1\over \pi}\im \Lambda(\br),\quad\Lambda(\br)=\Tr \hat M(\br)\mathcal G\hat V\mathcal G,\label{eq:ldos}
\end{align}
where the trace is taken over the internal degrees of freedom of the energy eigenstates $\nu$ and $\bk$. In Fourier space it is translated to
\begin{align}
N(\bq)={1\over 2\pi
i}\{\Lambda^*(-\bq)-\Lambda(\bq)\},\quad\Lambda(\bq)=\Tr\hat M(\bq)\mathcal G \hat V\mathcal G\label{eq:QPI}.
\end{align}
Substituting in \eqref{eq:QPI} the tip and impurity operators
\begin{align}
\hat M(\bq)
=\!\int_\bp\ket{\lambda;\bp}\hat m^\lambda_{\lambda'}\bra{\lambda';\bp-\bq},\quad\hat V\!=\!\int_{\bp,\bq} v(\bq,\br_0)\ket{\lambda;\bp}\hat v^\lambda_{\lambda'}\bra{\lambda';\bp-\bq},\end{align} 
as well as the diagonalized Green's function 
\eqref{ultimate_block}, we find that the QPI amplitude decomposes
in
\begin{align}\Lambda(\bq)=&\textstyle\sum_{\nu,\nu'\!,\bk,\bk'}\textstyle\sum_{\bG,\bG'\!,\bQ,\bQ'} v(\bq+\bQ+\bQ',\br_0)I^{\nu\bk}({\bG,\bG'})I^{\nu'\bk'}({\bG+\bQ,\bG'+\bQ'})\hat m^{\nu'\!-\gamma-\alpha~}_{\nu-\gamma}\hat v^{\nu-\gamma'\!-\alpha'}_{\nu'\!-\gamma'}\delta(\bq-\bk'+\bk+\bQ),\label{QPIexpandedSUP}\end{align}
with $\alpha={\bQ\cdot\be/ 2\pi}\!\!\!\mod \!n$ and $\alpha'={\bQ'\cdot\be/ 2\pi}\!\!\!\mod \!n$ which dictate whether $\bQ$ or $\bQ'$ crosses the Brillouin zone boundary along the nonsymmorphic direction $\be$. 
The summands in \eqref{QPIexpandedSUP} can be combined into
\begin{align}\Lambda(\bq)=\textstyle\sum_{\nu,\nu'\!,\bk,\bk'\!,\bQ,\gamma}\Lambda^{\nu\nu'}_{\bk\bk'\!,\gamma}(\bq,\bQ),\quad\Lambda^{\nu\nu'}_{\bk\bk'\!,\gamma}(\bq,\bQ)=\I^{\nu\nu'}_{\bk\bk'\!,\gamma}(\bq,\bQ)~\hat m^{\nu'\!-\gamma-\alpha~}_{\nu-\gamma}\delta(\bq-\bk'+\bk+\bQ),\end{align}
by carrying out the sum over the internal variables 
\begin{align}\I^{\nu\nu'}_{\bk\bk'\!,\gamma}(\bq,\bQ)=&\textstyle\sum_{\bG,\bG'\!,\bQ'} v(\bq+\bQ+\bQ',\br_0)\I^{\nu\bk}({\bG,\bG'})\I^{\nu'\bk'}({\bG+\bQ,\bG'+\bQ'})\hat v^{\nu-\gamma'\!-\alpha'}_{\nu'\!-\gamma'}\delta_{\omega^\gamma,e^{i\bG\cdot\be/n}}.\end{align}
Note that $\bQ$, and consequently $\alpha$, is fully determined by $\bq$ and the crystal momentum $\bk$ and $\bk'$, since the latter are only defined in the first Brillouin zone. On the other hand $\bQ'$, and consequently $\alpha'$, is summed over. This the a key ingredient to factor out the matrix elements of $\hat m$. Once we fix the relation,
\begin{align}
\hat m\hat R=\omega^\beta\hat R\hat m\end{align}
it implies that $\hat m_{\lambda'}^\lambda=0$ unless ${\omega^{\lambda-\lambda'}\!=\omega^\beta}$. This means that $\Lambda^{\nu\nu'}_{\bk\bk'\!,\gamma}(\bq,\bQ)$ vanishes unless ${\omega^{\nu-\nu'+\alpha}\!=\omega^\beta}$, for all $\gamma$. This is the selection rule presented in the main text.

We point out that the matrix elements introduced by $\hat v$ include $\alpha'$, not fixed by $\bq$. Therefore, they will not enter in the selection rule for $N(\bq)$ if the symmetry is nonsymmorphic. If, on the other hand, the symmetry flavor $\nu$ is symmorphic, we find that the matrix elements are independent of $\gamma$ or $\alpha$. That is, when we decompose the QPI amplitude we find the matrix elements to be independent of the Brillouin zone distance $\bQ$,
\begin{align}\Lambda(\bq)=\textstyle\sum_{\nu,\nu'\!,\bk,\bk'\!}\sum_{\bQ}\Lambda^{\nu\nu'}_{\bk\bk'\!}(\bq,\bQ),\quad\Lambda^{\nu\nu'}_{\bk\bk'\!}(\bq,\bQ)=\I^{\nu\nu'}_{\bk\bk'\!}(\bq,\bQ)~\hat m^{\nu'}_{\nu}\hat v_{\nu'}^\nu\delta(\bq-\bk'+\bk+\bQ),\end{align}
with
\begin{align}\I^{\nu\nu'}_{\bk\bk'}(\bq,\bQ)=&\textstyle\sum_{\bG,\bG'\!,\bQ'} v(\bq+\bQ+\bQ',\br_0)\I^{\nu\bk}({\bG,\bG'})\I^{\nu'\bk'}({\bG+\bQ,\bG'+\bQ'}).\end{align}
That is, the matrix elements of the impurity factor out and become as relevant as the tip matrix elements. Only for probing symmorphic symmetries the impurity and the tip have an interchangeable role. 

Finally, we note that the impurity can be invisible to given bands, and in this way further suppress QPI signals, both in the symmorphic and the nonsymmorphic cases. 

\section{Numerical calculations}

\begin{figure}[h!]
    \centering
    \includegraphics[width=0.9\columnwidth]{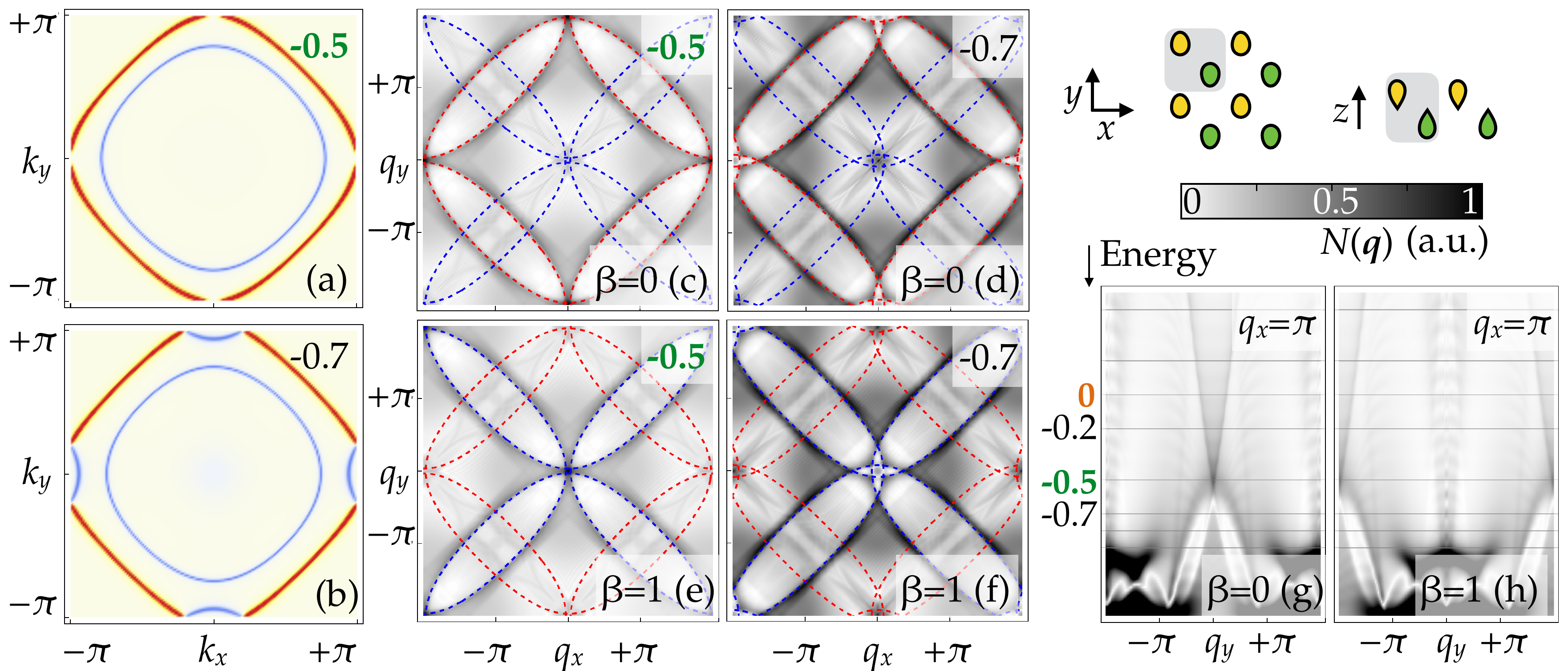}
    \caption{Additional numerical results for the two-dimensional model. (a,b) Density of states at energies $E=-0.5$ and $E=-0.7$. (c-f) QPI for two measurements with $\beta=0$ and $\beta=1$. (g-h) QPI energy cuts with $q_x=\pi$ where the dispersion of the QPI signal can be seen. With $\beta=0$ we see a crossing at $q_y=0$ while with $\beta=1$ the crossing appears at $q_y=\pi$.}
    \label{fig:my_label}
\end{figure}

In our numerical calculations, we use two one-dimensional models to represent the different phenomenology of a system invariant under a symmorphic and a nonsymmorphic symmetry, both include next nearest neighbour hopping. For a symmorphic chain, we consider
\begin{align}
H(k)=A_0+A_1\cos k+A_2\cos k~\sigma_z+A_3\sin (2k)+A_4\sin (2k)\sigma_z\label{sym},
\end{align}
with $A_i=(1,3,6,-1/4,3/4)$. Eq. \eqref{sym} commutes with $\sigma_z$ and explicitly breaks time-reversal symmetry. For a nonsymmorphic chain, we consider
\begin{align}
H(k)=A_0\sigma_z+A_1\cos k~\sigma_z+A_2\cos (k/ 2)\tau_z\sigma_z+A_3\cos (k/ 2)\tau_z+A_4\sin (k/ 2)\tau_z\sigma_z
\end{align}
with $A_i=(3,1,2,-1,-1)$. The nonsymmorphic symmetry is given by $e^{ik/2}\tau_z$ and $\tau_z$ defines the invariant $\nu$. This model preserves time-reversal symmetry, which pins the crossing of the two bands to the edge of the Brillouin zone.   
The above models are intended to represent a generic band structure compatible with a symmorphic and a nonsymmorphic symmetry. We do not discuss the microscopic origin of the different parameters $A_i$. 

In two dimensions we use the model introduced in Ref.\cite{Topp_Surface_2017} to describe the low energy theory of $\rm ZrSiS$, where the unit cell is composed by two sublattices, shifted by a translation $\bt=(\hat x+\hat y)/2$. In an embedded basis, it is given by 
\begin{align}
    H(\bk)=\mu+m\sigma_z+t^\pm_{xy}(\cos k_x+\cos k_y)(1\pm\sigma_z)+t\cos {k_x\over 2}\sin{k_y\over 2}\tau_x
\end{align} 
In the numerical simulations we have used $\mu=0$, $m=0.5$, $t_{xy}^-=-t_{xy}^+=0.5$, and $t=2.3$. It is invariant under the nonsymmorphic symmetry $e^{i(k_x+k_y)/2}\tau_x\sigma_z$. The eigenvalues of $\tau_x\sigma_z$ determine $\nu$. The nonsymmorphic symmetry and time-reversal symmetry guarantee that the bands cross at the boundaries of the Brillouin zone.

\section{Diagnosing Bloch symmetry with QPI: Application to ${\rm TaAs}$ }

\begin{figure}[h!]
    \centering
    \includegraphics[width=0.8\columnwidth]{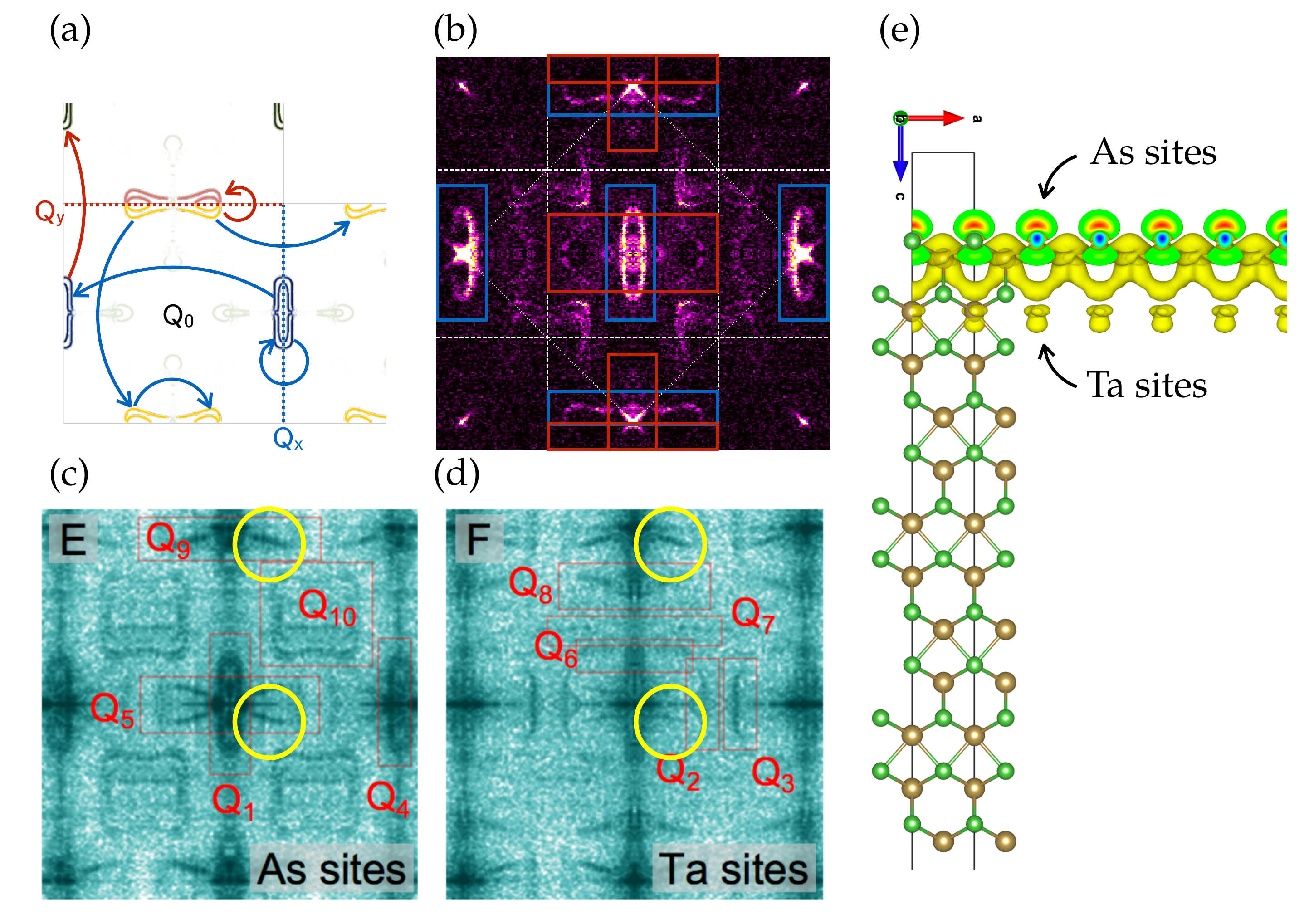}
    \caption{(a) Adapted from Ref.\cite{Batabyal_Visualizing_2016}, surface density of states of TaAs, where the ``bowtie" and the ``cigar" surface states are highlighted. Marked in blue and red arrows are the transitions with a strong and weak intensities in the QPI, respectively. We can identify the weak intensity signal to be associated with momentum transfer of $\bQ_y$. (b) From Ref. \cite{Batabyal_Visualizing_2016} QPI of surface states of TaAs, where the strong and weak signals are delimited in blue and red boxes, respectively (c-d) From Ref. \cite{Inoue_Quasiparticle_2016}, QPI with atomic resolution, where we have marked in yellow circles the QPI feature of our focus, associated with the ``bowtie" state. (c) QPI from As sites where the replications in the x and y direction do not show strong intensity modulations. (d) QPI from Ta sites, the replicated signal in $\hat y$ is suppressed, but not in $\hat x$. We can conclude that if this effect comes from matrix elements, it identifies a nonsymmorphic structure along the crystallographic direction $a$. (e) Density functional theory calculation of the charge density of the ``bowtie" surface state. We can distinctly see that contribution of the As sites is symmorphic, while the states around the Ta sites form a zigzag (nonsymmorphic) structure perpendicular to the surface. This calculation was performed by Binghai Yan and Huixia Fu.}
    \label{fig:my_label}
\end{figure}

\bibliographystyle{apsrev4-1}

\bibliography{readcube_export.bib}

\onecolumngrid

\end{document}